\documentstyle[aps,twocolumn,prl]{revtex}

\begin{document}
\draft
\preprint{}
%\widetext 
\title{    
Doping dependence of the resonance peak and incommensuration 
in high-$T_{c}$ superconductors
}
\author{
Jian-Xin Li$^{1}$ and Chang-De Gong$^{1,2}$}
\address{
1. National Laboratory of Solid States of Microstructure and Department of Physics, Nanjing University, Nanjing 210093, China\\
2. Chinese Center of Advanced Science and Technology (World Laboratory), P.O. Box 8730,\\ 
Beijing 100080, China}
\maketitle

\begin{abstract}
The doping and frequency evolutions of the incommensurate spin response and the resonance mode are studied based on the
scenario of the Fermi surface topology. We use the slave-boson mean-field approach to the $t-t^{\prime}-J$ model and including the antiferromagnetic fluctuation correction in the random-phase approximation. We find that the equality between the incommensurability and the hole concentration is reproduced at low frequencies in the underdoped regime. This equality
observed in experiments was explained {\it only} based on the stripe model before. We also obtain the downward dispersion
for the spin response and predict its doping dependence for further experimental testing, as well as a proportionality between the low-energy incommensurability and the resonance energy. Our results suggest a common origin for the incommensuration and the resonance peak based on the Fermi surface topology and the $d$-wave symmetry.   

\end{abstract}

%\newpage

%\section{INTRODUCTION}

The inelastic neutron scattering experiment plays an important role in the studies of the spin dynamics 
in high-$T_{c}$ superconductors. It can provide the momentum and frequency dependences of the dynamical
spin susceptibility. Over the past decade, the striking feature of the spin susceptibility observed in 
the momentum space is the incommensurate peak along the $(\pi,\pi+\delta \pi)$ direction~\cite{mason,yamada,lake}, 
and that in the frequency dependence is the resonance peak at the antiferromagnetic(AF) wavevector ${\bf Q}=(\pi,\pi)$~\cite{ros,fong}. These stimulate intensive experimental and theoretical studies. Recently, it 
is suggested that the presence of dynamic stripes is the origin of the observed incommensurate peak, at least for 
La$_{2-x}$Sr$_{x}$CuO$_{4}$~\cite{tra}. One of its stronge supports comes from the experiment by Yamada~\cite{yamada}
{\it et al.} which shows an equality between the incommensurability and 
the doping concentration (Yamada plot) in the underdoped La$_{2-x}$Sr$_{x}$CuO$_{4}$. This equality follows naturally 
from the static stripe model, but has not been explained in any other way up to now~\cite{millis}. However, it is now 
unclear how to explain the resonance peak based on this scenario. On the other hand, the gross features of the incommensurate spin fluctuations can also be explained based on the scenario of the nested Fermi surface(FS)~\cite{bri,voo}
and the resonance peak is thought to be a collective spin excitation mode in this framework~\cite{blu,bri}. Obviously,
more fine and detailed experimental data are helpful for selecting out or ruling out the above models. Other than 
the Yamada plot, we note that some new experimental developments have also been reported. Among these are the  
detailed evolutions of the resonance peak and incommensurability with doping~\cite{fong2,dai}, and the dispersion 
connecting these two structures which has a downward curvature opposite in sign to a conventional magnon dispersion~\cite{arai,bourges}. 

In view of these new experimental observations, we present in this paper a detailed investigation of the doping
and frequency dependences of the incommensurability and the resonance peak energy, and their relationship based on the nested Fermi surface scenario. We start from the slave-boson approach to the two-dimensional(2D) $t-t^{\prime}-J$ model. 
The aim of choosing this approach is that we can incoprate the evolutions of the FS and the superconducting(SC) gap with 
dopings self consistently at the mean-field level and determines their values according to the well accepted input
parameters $t,t^{\prime},J$. This enables us to carry out a quantitative study of the doping and frequency evolutions of  spin fluctuations. In most previous studies~\cite{voo,blu,nor}, a tight-binding dispersion and the magnitude of the SC gap which are inferred from related experimental data for a fixed doping concentration are taken. Consequently, these investigations are limited to a fixed hole doping. Our main results are: 1) We find that the equality between the low-energy incommensurability and doping density in the underdoped regime is reproduced. This may provide an alternative to its exclusive stripe-phase explanation. 2) The downward dispersion of the collective mode is reproduced and understood as arising from the Fermi surface geometry, and its doping dependence is presented to be compared with further experiments. 
3) Below a crossover frequency, the incommensurate peak shifts from the $(\pi,\pi+\delta \pi)$ direction to the diagonal direction due to the node-to-node excitations. Above the resonance frequency, the incommensurate peak occurs again. It is along the $(\pi,\pi+\delta \pi)$ direction in the underdoped regime $(x \le 0.13)$ and along the diagonal direction 
in the optimally doped and overdoped regimes $x \ge 0.14$. 4) The resonance energy has a linear relation with doping in the underdoped regime and saturates near optimal doping, then slightly decreases with doping in the overdoped regime. As a result, a proportionality between the low-energy incommensurability and the resonance energy is obtained. Most of these results are consistent with experimental data~\cite{dai,he} and the others wait for further experimental testing. Therefore our result presents an alternative explanation of the incommensuration and resonance peak in the spin response based on the $d$-wave superconductor with tight-binding dispersion. However, the abrupt saturation of the incommensurability near the optimal doping and in the overdoped regime which is observed both in La$_{2-x}$Sr$_{x}$CuO$_{4}$~\cite{yamada} and YBa$_{2}$Cu$_{3}$O$_{6+x}$~\cite{dai} is not found. Possible reason for this is discussed. 

In the slave-boson approach to the $t-t^{\prime}-J$ model~\cite{bri,ubb}, the physical electron operators $c_{i\sigma}$
are expressed by slave bosons $b_{i}$ carrying the charge and fermions
$f_{i\sigma}$ representing the spin; $c_{i\sigma}=b_{i}^{+}f_{i\sigma}$. In the SC state,
we consider the order parameters $\Delta_{ij}=<f_{i\uparrow}f_{j\downarrow}-f_{i\downarrow}f_{j\uparrow}>=\pm \Delta_{0}$ with the $d$-wave symmetry and $\chi_{ij}=\sum_{\sigma} <f_{i\sigma}^{+}f_{j\sigma}>=\chi_{0}$, in which bosons condense $b_{i}\rightarrow <b_{i}>=\sqrt x$ ($x$ is the hole concentration). Then, the slave-boson mean-field Hamiltonian of the 2D $t-t^{\prime}-J$ model in the SC state is,
\begin{eqnarray} 
H_{m}& =& \sum_{k\sigma}\epsilon_{k}f^{\dag}_{k\sigma}f_{k\sigma}
-\sum_{k}\Delta_{k} (f^{\dag}_{k\uparrow}f^{\dag}_{-k\downarrow}+h.c.) \\ \nonumber
& & +2NJ'(\chi_{0}^{2}+\Delta_{0}^{2}),
\end{eqnarray}
where $\epsilon_{k}=-2(x t+J'\chi_{0})[\cos(k_{x})+\cos(k_{y})]-
4x t'\cos(k_{x})\cos(k_{y})-\mu$ is the dispersion for fermions,
and $\Delta_{k}=2J'\Delta_{0}[\cos(k_{x})-\cos(k_{y})]$, with
$J'=3J/8$. The normal state Fermi surface at doping $x=0.12$ for this dispersion is shown in Fig.1. The mean-field parameters $\chi_{0}$, $\Delta_{0}$ and the chemical potential $\mu$ for different dopings are obtained from a self-consistent calculation~\cite{ubb}.

It has been shown~\cite{bri,li,li2} that the inclusion of the AF fluctuation correction is necessary to account for some
of the spin and charge dynamics. Formally, this can be done by perturbing around the mean-field Hamiltonian, i.e., we write 
$H=H_m+H^{\prime }$ with $H$ the usual 2D $t-t^{\prime}-J$ Hamiltonian, and treat $H^{\prime }$ as a perturbation. In principle, all fluctuations are included.
However, different selection of subset diagrams may result in different kinds of fluctuations.
For the spin fluctuation, the usual random phase approximation(RPA) selects a series of ring diagrams as 
shown in Ref.~\cite{bri,li,li2}. Then, the renormalized spin susceptibility is given by,
\begin{equation}
\chi({\bf q},\omega)={ \chi_{0}({\bf q},\omega) \over {1+\eta J\gamma_{q}
\chi_{0}({\bf q},\omega)}}.
\end{equation}
where $\gamma_{q}=\cos (q_{x})+\cos (q_{y})$. $\chi_{0}$ is the bare spin susceptibility which comes from the fermionic bubble and is given by the usual BCS form~\cite{schrieffer}. As done before~\cite{bri,li,li2}, we choose $\eta=0.34$ instead of $\eta=1$ in order to set the AF instability at $x=0.02$ which is the experimental observed value.

Numerical calculations are performed by dividing the Brillouin zone into 1024$\times$1024 lattices, with $t=2J,t^{\prime }=-0.45t$ and $J=0.13$eV. The damping rate of the fermionic quasiparticles is simulated to be $\Gamma =0.004J$, however
we note that the incommensurability and the resonance energy is not subjected to the change of $\Gamma$ when it is 
below $0.1J$. Therefore, the only adjusted parameter throughout the paper is $\eta$ which has been fixed above.

In Fig.2, the peak position of the imaginary part of the renormailized spin susceptibility is shown as a function of frequency, for doping concentrations $x=0.08,0.14$ and $0.20$ which may
correspond to the underdoped, optimally doped and overdoped regimes. At their resonance energies ($0.375J, 
0.53J$ and $0.545J$,respectively), the $q$ peaks are at $\delta=0$, i.e. at $(\pi,\pi)$ point. When the frequency is 
reduced from the resonance energy, the incommensurate peak occurs and is found to be in the $(\pi,\pi+\delta \pi)$
direction. In particular, a downward curvature for the peak dispersion is observed for various dopings. This is in 
agreement with 
experiments~\cite{arai,bourges}. To understand its origin, we show the frequency dependences of the bare spin susceptiblity Im$\chi_{0}$ at $x=0.14$ for different wavevectors ${\bf q}=(\pi,0.72\pi),(\pi,0.8\pi)$ and $(\pi,0.9\pi)$ in Figs.3(a) and 3(b), and at ${\bf q}=(\pi,0.84\pi)$ for different dopings $x=0.08, 0.14$ and 0.20 in Figs.3(c) and 3(d). Firstly, it is important to point out that the bare spin susceptiblity Im$\chi_{0}$ is {\it incommensurate} for all energies and dopings in the ranges considered here. Moreover, its peak is along the $(\pi+\delta \pi, \pi)$ or $(\pi, \pi+\delta\pi)$ directions. It comes from the FS nesting effect
as has been shown by Brinckmann and Lee~\cite{bri}. In the current framework, the origin of the resonance peak is ascribed to a collective spin excitation corresponding to the pole of the renormalized spin susceptibility, i.e. $1+\eta J\gamma_{\bf Q}{\rm Re}\chi_{0}({\bf Q},\omega )=0$, and negligibly small Im$\chi _{0}({\bf Q},\omega )$~\cite{bri,blu,li}. At the commensurate wavevector ${\bf Q}=(\pi,\pi)$, there is a step-like rise in Im$\chi_{0}$ (shown as the soild line in Fig.3(b)) at the threshold energy for particle-hole(p-h) excitations. Due to this step-like rise, a logarithmic singularity in Re$\chi_{0}$ occurs via the Kramers-Kroenig relation. This singularity shifts downward the collective mode energy and leads it to situate in the spin gap, so no damping is expected for the mode. For the energy band $\epsilon_{k}$ and the SC gap considered here, the initial and final states of the p-h excitations at the threshold energy (minmum excitation energy) lie on the Fermi surface. So, the threshold excitation with transition wavevector ${\bf Q}$ corresponds to the excitation $A\rightarrow B$ as schmatically shown in Fig.1. The step-like rise is caused mainly by
the van Hove(vH) singularity around $(\pi,0)$ and the $d$-wave symmetry of the SC gap~\cite{blu,li,li2}. In this case, the collective spin excitation at $(\pi,\pi)$ which is determined by the pole condition in the RPA correction $1+\eta J\gamma_{q}\chi_{0}$ dominates the structure of the renormalized spin susceptibility Im$\chi$ and shows up as a strong commensurate resonance peak, although the bare spin susceptibility Im$\chi_{0}$ is incommensurate. When the wavevector
${\bf q}$ moves away from ${\bf Q}$, the p-h transitions will move to the node direction and that connecting $(0,\pi)$ and $(\pi,0)$, respectively. As a result, the single step-like rise splits into two as shown in Fig.3(b). One is at the energy below it and the other above it, which correspond to the excitations $E\rightarrow F$ and $G\rightarrow H$ in Fig.1, respectively. The pole equation still satisfies around the high-energy peak (The line denoting $-1/\eta J\gamma_{q}$ for ${\bf q}\ne {\bf Q}$ will rise above the thin line in Fig.3(a) due to $|\gamma_{q}|<|\gamma_{Q}|$), but the pole position falls well above the gap and the resultant p-h excitations are overdamped. On the other hand, though the peak caused by the low-energy jump does not satisfy the pole equation, because its jump height is reduced due to the weakening of the vH singularity at the ${\bf q}$ points $E$ and $F$ which are away from $(0,-\pi)$ and $(\pi,0)$, it still affects the structure of the spin response via the RPA correction because of the small damping of the excitations. We can see from Fig.3(b) that, when the deviation of the wavevector ${\bf q}$ from ${\bf Q}$ increases, the corresponding low-energy step-like rise decreases. From Ref.~\cite{bri}, we also know that the incommensurability of Im$\chi_{0}$ increases with the decrease of frequency. So, the peak dispersion determined both from the RPA correction and the bare spin susceptibility
Im$\chi_{0}$ has a downward curvature. But, which one mainly determines the position of the incommensurate peak in the renormalized spin susceptibility is interesting. This issue is related to the argument which attributes the incommensurate magnetic response and the commensurate resonance to belong to parts of the same collective mode~\cite{bourges,bat}. In the previous investigation~\cite{bri}, the incommensurate magnetic response is explained solely based on the structure of Im$\chi_{0}$, because of the consideration $\chi \approx \chi_{0}$. However, we find that the peak position of Im$\chi$ in $q$ space, {\it i.e.} the incommensurability, is mainly determined by the RPA correction
$1+\eta J\gamma_{q}\chi_{0}$ and in turn by the low-energy step-like rise. This can be seen 
clearly from Fig.3(b), where the dashed-dotted line, the dotted line and the dashed line denote the imaginary parts of the bare spin susceptibility Im$\chi_{0}$ at ${\bf q}=(\pi, 0.72\pi), (\pi, 0.8\pi)$ and $(\pi, 0.9\pi)$, respectively. These wavevectors are the incommensurate peak positions in the renormalized spin susceptibility Im$\chi$ at frequencies $\omega=0.15J, 0.3J$ and $0.45J$, respectively. Clearly, the low-energy step-like rises in Im$\chi_{0}$ shown in Fig.3(b) and consequently the low-energy peaks in their real parts shown in Fig.3(a) are at the same values as the above frequencies for the corresponding wavevectors. On the other hand, the incommensurabilities in Im$\chi_{0}$ for $\omega=0.15J, 0.3J$ and $0.45J$ are at ${\bf q}=(\pi, 0.72\pi), (\pi, 0.76\pi)$ and $(\pi, 0.88\pi)$. The same situation is found for other doping concentrations. Therefore, both the resonance peak and the incommensurability are mainly determined by the step-like rise of Im$\chi_{0}$. 

Below the lowest frequency for each doping density shown in Fig.2, we find that a crossover of the incommensurate peaks 
from the $(\pi, \pi+\delta\pi)$ direction to the diagonal direction occurs. It is because the excitations will be limited 
to be along the diagonal direction(the node-to-node excitation), such as $E \rightarrow F$ shown in Fig.1, when the excitation energy is below the crossover energy due to the energy conservation law and the $d$-wave symmetry of the SC gap. It is a specular feature for the model based on the FS topology and may also act to distinguish between the stripe and the FS nesting models. Unfortunately, the crossover energy for the underdoped system is too low to be available for the experimental observation due to the dim scattering intensity~\cite{dai}. However, we can see from Fig.2 that it increases with doping density and is about $0.135J, 0.175J$ and $0.2J$ for the doping $x=0.08, 0.14$ and 0.20, respectively. So, we expect that it may be in the experimental observable range for the overdoped cuprates. For example, the crossover energy 
for $x=0.20$ is $0.2J\approx 26$meV which is experimentally accessible according to the recent report~\cite{dai}. Above the resonance frequency, the incommensurate peak occurs again. In particular, we find that only in the underdoped regime $(x \le 0.13)$ are the incommensurate peaks along the $(\pi,\pi+\delta \pi)$ direction. They are along the diagonal direction in the optimally doped and overdoped regimes $(x \ge 0.14)$. It is in fact reflects the doping dependence of the intensity ratio of the incommensurate peak along the $(\pi,\pi+\delta \pi)$ direction to that along the diagonal direction, which decreases with the increase of doping density when the frequency is above the resonance
energy~\cite{note}. From Fig.2, one can see that the incommensurability above the resonance frequency increases with frequency, so the dispersion shows a upward curvature which is similar to that of a massive magnon in a disorder N\'{e}el state.

We show the doping dependences of the incommensurability $\delta$ in the $(\pi,\pi +\delta \pi)$ direction in Fig.4(a) for energies $\omega=0.15J, 0.175J, 0.20J, 0.25J$ and $0.35J$. In the underdoped regime, the incommensurability increases with the increase of doping and exhibits a nearly linear doping dependence for all energies. In the high-energy range, the incommensurability decreases with frequency. For La$_{2-x}$Sr$_{x}$CuO$_{4}$, it is found that the incommensurability is energy independent for $\hbar\omega<15$meV. The same behavior is found here when the frequency is reduced to be below about $\omega=0.20J$, as can be seen from Fig.4(a). By a closer inspection of the doping dependence of the incommensurability at low frequencies such as $\omega=0.175J$ and $0.15J$, we find that the equality $\delta =2x$ holds in the doping range from $x=0.06$ to $x=0.13$. We note that the incommensurability $\delta$ defined here is twice that used by experimentalists which is indexed in units of the reciproal lattice vectors. This equality has been observed at low frequencies in underdoped cuprates~\cite{yamada,dai} and is believed to be explained only in the stripe model before~\cite{millis}. Therefore, our result provides an alternative explanation based on the FS nesting. However, 
the incommensurability increases continually in the whole doping range, which is different from the experimental
observation that it saturates near the optimal doping and in the overdoped regime~\cite{yamada,dai}. 
We also show the doping dependences of the resonance energy $E_{r}$ in Fig.4(b). This result has been reported by one of us (Li) and his coworkers in a smaller doping range before~\cite{li}. Similarly, a nearly linear doping dependence is found in the underdoped regime, then it saturates in the slightly overdoped regime and eventually decreases with the further increase of dopings. Due to the similar doping dependence in the underdoped regime, the ratio of $E_{r}$ to $\delta$ shows a linear behavior as can be seen in the inset of Fig.4(b). These results are remarkably in agreement with experiments~\cite{dai,he}.
To understand these behaviors, we show the frequency dependences of the bare spin susceptibility $\chi_{0}$ at the incommensurate ${\bf q}$-point $(\pi, 0.84\pi)$ for different dopings in Figs.3(c) and 3(d), and at the commensurate ${\bf q}$-point $(\pi,\pi)$ in the inset of Fig.3(d). We find that the step-like rise at $(\pi,\pi)$ and the low-energy step-like rise at $(\pi, 0.84\pi)$ increases monotonously with doping in the underdoped regime. But the former saturates in the optimally doped regime and then decreases with the further increase of dopings, and the latter increases with dopings in the whole doping range considered here. According to the above discussion, it is this feature that leads to the special doping dependences for both cases. From this reasoning, we may also argue that the saturation of the incommensurability observed experimentally near and after the optimal doping 
should not be related {\it only} to the change of the FS topology. Whether new physics such as the stripe phase is required or some additional factors such as the change of the quasiparticle nature due to a quantum critical point near the optimal doping~\cite{sachdev} should be taken into consideration is a open question.         

In summary, we have examined the doping and frequency dependences of the incommensurability and the resonance energy based on the Fermi surface topology. The calculations are carried out by use of the slave-boson mean-field approach to the $t-t^{\prime}-J$ model and including the antiferromagnetic fluctuation correction in the random-phase approximation. We find that the equality of the incommensurability and the doping density exists at lower frequencies in the underdoped regime, which has not been explained alternatively other than the stripe model before. The downward dispersion is reproduced and its doping dependence is presented for further experimental testing. We also find a good linear behavior between the incommensurability and the resonance energy in the underdoped regime, which is consistent with experiments. Our results may suggest a common origin for the incommensurate spin response and the resonance peak based on the
Fermi surface topology and the $d$-wave superconductivity.

%\section*{ACKNOWLEDGMENTS}
JXL thank T.K.Lee and C.Y.Mou for useful conversations. He was supported by the National Nature Science Foundation of
China. CDG thanks the support by the Ministry of Science and Technology of China(NKBRSF-G19990646).

\newpage
\section*{FIGURE CAPTIONS}
%\vspace{0.2cm}
Fig.1  Fermi surface for the dispersion $\epsilon_{k}$ (thick solid line) at hole doping $x=0.12$.
The thin solid and dashed lines with arrow denote the threshold particle-hole excitations for different wavevectors(see text).

%\vspace{0.2cm}
Fig.2 Peak positions of the imaginary part of the renormalized spin susceptibility in the $q$ space as a 
function of frequency for doping concentrations $x=0.08,0.14$ and 0.20. In the downward dispersion range, 
the incommensurate peaks are along the $(\pi,\pi+\delta \pi)$ direction, so the horizontal axis is in unit of
$(1,q)\pi$. Below $\omega=0.135J, 0.175J$ and $0.2J$ at dopings $x=0.08, 0.14$ and 0.20, respectively, and for frequencies above the resonance energy at $x=0.14$ and 0.20, they are along the diagonal direction. In the latter case,
the horizontal axis is in unit of $(q,q)\pi$. 

%\vspace{0.2cm}
Fig.3 Frequency dependence of the bare spin susceptibility $\chi_{0}$. Figures (a) and (b) are the real
and imaginary parts of the bare spin susceptibility at doping $x$=0.14 for different ${\bf q}$. The dashed-dotted 
line, the dotted line and the dashed line denote spin susceptibilities at ${\bf q}=(\pi, 0.72\pi), (\pi, 0.8\pi)$ 
and $(\pi, 0.9\pi)$, respectively. The thin line in figure (a) denotes $-1/2\eta J$, so its cross with the real part corresponds to the pole of the renormalized spin susceptibility. Figures (c) and (d)
are the real and imaginary parts of the bare spin susceptibility at ${\bf q}=(\pi,0.84\pi)$ for different 
dopings. The solid line, dashed line and dotted line denote spin susceptibility at $x=0.08, 0.14$ and $0.2$, respectively.  
The inset of figure (d) shows the frequency dependence of Im$\chi_{0}$ at ${\bf q}=(\pi,\pi)$ for the doping
$x=0.08$(solid line),0.14(dashed line),0.20(dotted line) and 0.24(dashed-dotted line). 

%\vspace{0.2cm}
Fig.4 (a) Doping dependences of the incommensurability $\delta$ in the $(\pi,\pi+\delta \pi)$ direction 
for different frequencies. (b) Doping dependence of the resonance energy $E_{r}$. The inset of figure (b) shows the 
ratio between the incommensurability and the resonance energy. The solid line in the inset is a guide to the eye. 
Note that the incommensurability $\delta$ defined here is twice that used by experimentists which is indexed in units 
of the reciproal lattice vectors.

\begin{thebibliography}{99}
\bibitem{mason} T.E. Mason, G. Aeppli, and H.A. Mook, Phys. Rev. Lett. {\bf 68}, 1414 (1992).
\bibitem{yamada} K. Yamada, C.H. Lee, K. Kurahashi, J. Wada, S. Wakimoto, S. Ueki, H. Kimura,
Y. Endoh, S. Hosoya, G. Shirane, R.J. Birgeneau, M. Greven, M.A. Kaster, and Y.J. Kim, 
 Phys. Rev. B {\bf 57}, 6165 (1998).
\bibitem{lake} B. Lake, G. Aeppli, T.E. Mason, A. Schroder, D.F. McMorrow, K. Lefmann, M. Isshiki,
M. Nohara, H. Takagi, and S.M. Hayden, Nature (London) {\bf 400}, 43 (1999).
\bibitem{ros} J. Rossat-Mignod, L.P. Regnault, C. Vettier, P. Bourges, P. Burlet, J. Bossy, J.Y. Henry,
and G. Lapertot,  Physica C {\bf 185-189}, 86 (1991). 
\bibitem{fong} H.F. Fong, B.Keimer, P.W. Anderson, D. Reznik, F. Do\v{g}an, and I.A. Aksay, 
Phys. Rev. Lett. {\bf 75}, 316 (1995);
\bibitem{tra} J.M. Tranquada, B.J. Sternlieb, J.D. Axe, Y. Nakamura, and S. Uchida, 
Nature (London) {\bf 375}, 561 (1995); J.M. Tranquada, J.D. Axe, N. Ichikawa, Y. Nakamura,
S. Uchida, and B. Nachumi, Phys. Rev. B {\bf 54}, 7489 (1996).
\bibitem{millis} J. Orenstein and A.J. Millis,  Science {\bf 288}, 468 (2000).
\bibitem{bri} J. Brinckmann and P.A. Lee,  Phys. Rev. Lett. {\bf 82}, 2915 (1999).
\bibitem{voo} K.K. Voo, W.C. Wu, J.X. Li and T.K. Lee,  Phys. Rev. B {\bf 61}, 9095 (2000);
 D.K. Morr and D. Pines, Phys. Rev. B {\bf 61}, 6483 (2000); Y.J. Kao, Q. Si, and K. Levin, 
Phys. Rev. B {\bf 61}, 11898(2000).
\bibitem{blu} G. Blumberg, B.P. Stojkovic and M.V. Klein,  Phys. Rev. B {\bf 52}, 15 741 (1995);
D.Z. liu, Y. Zha and K. Levin, Phys. Rev. Lett. {\bf 75}, 4130 (1995); J.X. Li, W.G. Yin, and 
C.D. Gong,  Phys. Rev. B {\bf 58}, 2895 (1998).
\bibitem{fong2} H.F. Fong, P. Bourges, Y. Sidis, L.P. Regnault, J. Bossy, A. Ivanov, D.L. Milius,
I.A. Aksay, and B. Keimer,   Phys. Rev. B {\bf 61}, 14 773 (2000).
\bibitem{dai} P. Dai, H.A. Mook, R.D. Hunt, and F. Do{\'g}an, cond-mat/0011019 (unpublished).
\bibitem{arai} M. Arai, T. Nishijima, Y. Endoh, T. Egami, S. Tajima, K. Tomimoto, Y. Shiohara, 
M. Takahashi, A. Garret, and S.M. Bennington,  Phys. Rev. Lett. {\bf 83}, 608 (1999).
\bibitem{bourges} P. Bourges, Y. Sidis, H.F. Fong, L.P. Regnault, J. Bossy, A. Ivanov, and B. Keimer,
 Science {\bf 288}, 1234 (2000).
\bibitem{nor} M.R. Norman, cond-mat/0010298 (unpublished).
\bibitem{he} H. He, Y. Sidis, P. Bourges, G.D. Gu, A. Ivanov, N. Koshizuka, B. Liang, C.T. Lin, 
L.P. Regnault, E. Schoenherr, and B. Keimer, cond-mat/0002013 (unpublished).
\bibitem{ubb} M.U.Ubbens and P.A.Lee,  Phys. Rev. B {\bf 46}, 8434 (1992); T. Tanamoto, K. Kuboki and H. Fukuyama,
J. Phys. Soc. Jpn. {\bf 60}, 3072 (1991).
\bibitem{li} J.X. Li, C.Y. Mou, and T.K. Lee,  Phys. Rev. B {\bf 62}, 640 (2000).
\bibitem{li2} J.X. Li, C.Y. Mou, C.D. Gong and T.K. Lee, cond-mat/0102448 (unpublished).
\bibitem{schrieffer}  J. R. Schriffer, Theory of Superconductivity,
(W.A.Benjamin, Massachusetts, 1964).
\bibitem{note} We note that this intensity ratio nearly does not change with the variation of dopings in the downward
dispersion range.
\bibitem{bat} C. D. Batista, G. Ortiz, A. V. Balatsky, cond-mat/0008345 (unpublished).
\bibitem{sachdev} S. Sachdev, Science {\bf 288}, 5465 (2000).

\end{thebibliography}
\end{document}